# Ultrahigh-speed optical coherence tomography based on a 100 MHz and 100 nm swept source

Dongmei Huang, Feng Li, *Senior Member, OSA, Member, IEEE,* Zihao Cheng, Xinhuan Feng, and P. K. A. Wai, *Fellow, OSA, Fellow, IEEE*

*Abstract*—We demonstrate an ultrahigh-speed optical coherence tomography (OCT) based on a 100 MHz swept source (SS). An all polarization-maintaining figure-9 mode-locked fiber laser is used as the seed laser. After nonlinear spectral expansion in an Erbium-doped fiber amplifier, a flat top spectrum with respectively 1-dB and 10-dB bandwidths of 73.7 nm and 106 nm is obtained. The broadband femtosecond pulse is time stretched to a swept signal in a section of dispersion compensation fiber with a total dispersion of –84 ps/nm. With the swept source, the axial resolution of the SS-OCT is measured to be 21 μm with a 6 dB sensitivity roll-off length of 3 mm. A tomographic image of an encoding disk and a hard disk jointly rotating at 17,000 rpm was acquired by using the SS-OCT with a high imaging quality.

## I. INTRODUCTION

OPTICAL coherence tomography (OCT) is a powerful technique for biological imaging and industrial inspection [1]. The development of OCT technology has undergone three generations as time domain OCT (TD-OCT), spectral domain OCT (SD-OCT) and swept source OCT (SS-OCT)[2]–[4]. Particularly, SS-OCT is the most promising and practical technology of the next generation OCT since it avoids the mechanical movement of the reference arm in TD-OCT and overcomes the fundamental acquisition speed limitation of spectrometers such as CCD or CMOS in SD-OCT. Swept source OCT [1], [5], [6] can simultaneously achieve MHz A-scan rates and high resolutions by adopting high-speed photodetectors and broadband swept sources. The imaging speed and axial resolution of SS-OCT is mainly determined by the swept source [4], [7]. Novel techniques have been developed to build high speed swept sources for SS-OCT, including micro-electro-mechanical system (MEMS) vertical-cavity surface-emitting lasers (VCSELs) [8] and Fourier domain mode-locked fiber lasers [4], [6]. However, the fundamental sweep rates of these schemes are limited to sub-MHz level by the inertia-restricted wavelength tuning elements. Ultrafast swept sources have been constructed by inertia free time stretching of high repetition rate broadband pulses in dispersive elements such as dispersion compensation fibers (DCFs) or chirped fiber Bragg gratings (CFBGs). Stretched pulse mode-locking (SPML) with a pair of time stretching elements with respective positive and negative dispersion in the cavity could generate swept signals with MHz sweep rates by active harmonic modulation [9]–[11]. In contrast, time stretching outside cavity is more flexible and it has demonstrated sweep rates up to tens of megahertz [12]–[15]. For optimal performance of the SS-OCT, the seed laser used for time stretching should be highly stable and have a high repetition rate, as it is directly the sweep rate of swept signals. The spectrum should also be coherent, flat, and broadband.

In the first demonstration of time stretched SS-OCT, a filtered supercontinuum (SC) with a repetition rate of 5 MHz is used as the seed [12]. Since it is currently impossible to fabricate an optical fiber with high monotonic dispersion and a low loss in such a broad spectral range, the SC should be filtered before time stretching [12], [13], [16], which wastes a large portion of the power. The large shot to shot fluctuation of the spectrum further degrades the performance of time stretched SC[17], [18]. Stable broadband spectra with high repetition rates are available in mode-locked fiber lasers (MLFLs) based on nonlinear polarization evolution (NPE) [19]–[21] or nonlinear amplified loop mirror (NALM) [22], [23]. In compact NPE cavities, a bandwidth of 135 nm [19] and a repetition rate of 517 MHz [21] have been demonstrated, respectively. A dispersion-managed NPE MLFL with a 102 nm 10-dB bandwidth and a 44.5 MHz repetition rate has been adopted in time stretching SS-OCT, which is the current A-scan rate record of practical SS-OCT with a bandwidth of ~100 nm [24]. The major challenge to build robust equipment with NPE MLFLs is their sensitivity to environmental perturbations [25], which could be circumvented in all polarization-maintaining (PM) NALM fiber lasers [26]. However, the lasing bandwidth of NALM lasers is typically narrower than that of NPE lasers and self-starting is challenging in most figure-8 configurations.

In this paper, we will demonstrate a stable, coherent, and flat-spectrum time stretched sweep source at 100 MHz sweep rate and a 100 nm effective sweep range, which is defined as the 10-dB bandwidth of the spectrum. A robust figure-9 all PM MLFL is used as the seed. The flat and broad spectrum is generated in a subsequent polarization-maintaining Erbium-doped fiber amplifier (PM-EDFA). The ultrafast swept source is successfully adopted in SS-OCT and is characterized by measuring the sensitivity roll-off and imaging of high-speed rotating disks.

This work was supported in part by National Key R&D Program of China under grant 2019YFB1803900; in part by Shenzhen Science and Technology Innovation Commission under grant SGDX2019081623060558; in part by Research Grant Council of Hong Kong SAR under grants PolyU152471/16E and PolyU152241/18E; in part by The Hong Kong Polytechnic University under grants 1-BBAJ and 1-ZVGB. *(Corresponding author: Feng Li).*

Dongmei Huang, Feng Li, Zihao Cheng, and P. K. A. Wai are with the Photonics Research Centre, Department of Electronic and Information Engineering, The Hong Kong Polytechnic University, Hung Hom, Hong Kong SAR, China, and The Hong Kong Polytechnic University Shenzhen Research Institute, Shenzhen 518057, China (e-mail: dongmei.huang@connect.polyu.hk; enlf@polyu.edu.hk; zihao.cheng@connect.polyu.hk; alex.wai@polyu.edu.hk).

Xinhuan Feng is with the Guangdong Provincial Key Laboratory of Optical Fiber Sensing and Communications, Institute of Photonics Technology, Jinan University, Guangzhou 510632, China (e-mail: tfengxh@jnu.edu.cn).

## II. EXPERIMENTAL SETUP

### A. Experimental setup of 100 MHz swept source

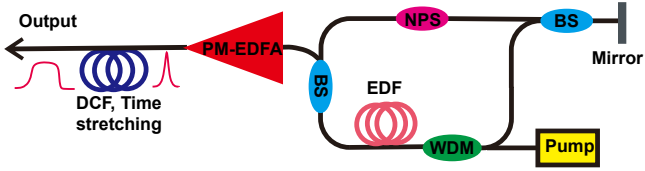

Fig. 1. Experimental setup of the swept source based on time stretching of pulses from a mode-locked fiber laser. WDM: wavelength division multiplexer, EDF: Erbium-doped fiber, OC: optical coupler, BS: beam splitter, NPS: nonreciprocal phase shifter, PM-EDFA: Polarization-maintaining Erbium-doped fiber amplifier, DCF: dispersion compensation fiber.

The experimental setup of the swept source is shown in Fig. 1. A custom-built figure-9 MLFL with a dispersion managed PM-EDFA (Menlo System, C-Fiber) is used as the seed of the whole system [23]. In the experiment, the fiber tail is modified and the pump powers of the seed laser and the PM-EDFA are carefully tuned to obtain a flat spectrum. The MLFL uses a compact all PM NALM cavity with a nonreciprocal phase shifter to ensure self-starting [23], [26]. In operation, the mode-locking can be self-started by simply tuning the pump power. Following the PM-EDFA, a section of DCF (YOSC, DCM-652-05) with a nominal dispersion of –84 ps/nm at 1545 nm and an insertion loss of 1.3 dB is used as the dispersion element to stretch the broadband pulse. To avoid strong nonlinear effects in the DCF, we use a variable optical attenuator to adjust the input power of the DCF, which could be replaced by either a power splitter for parallel systems or a section of large mode area fiber to preliminarily chirp the pulse into a picosecond pulse to lower the peak power.

### B. Point spread function measurement and OCT system

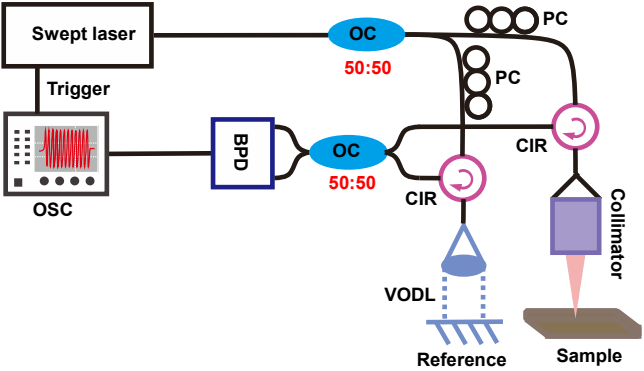

Fig. 2. A schematic diagram of the OCT system. OSC: oscilloscope, BPD: balanced photodetector, CIR: circulator, OC: optical coupler, VODL: variable optical delay line, PC: polarization controller.

The high speed swept source is then deployed in an OCT system with the schematic shown in Fig. 2. The input power is split into the two arms using a 50:50 optical coupler. The reference arm of the Michelson interferometer is a motorized reflective variable optical delay line (VODL). The sample beam is focused onto the sample with a collimator. The circulators in the two arms route the reflected signals into another 50:50 coupler to be re-combined. The combined signals are detected by a 43 GHz balanced photodetector (BPD, Finisar, BPDV2120R) to remove the DC background. A high-speed real-time oscilloscope (Tektronix, DPO75902SX ATI) is used in the acquisition of the electrical signal.

## III. EXPERIMENTAL RESULTS AND DISCUSSIONS

### A. Generation of the 100 MHz broadband swept source

The pulse spectrum from a laser is limited by the bandwidth of the gain element inside a laser cavity. Nonlinear spectrum broadening outside the laser cavity could be used to broaden the pulse spectrum. Spectral broadening in fiber amplifiers has been proven as an effective method to obtain high-quality broadband spectra [27]–[30]. It is known that the propagation of a higher-order soliton will evolve to a double hump spectrum with suppression of the central part. Such higher order soliton evolution is also an important process in supercontinuum generation before soliton fission. During the spectral evolution from a single to a double hump profile, there is a short period that the spectrum has a flat-top profile. To have a clear understanding of the evolution to generate a flat-top spectrum, we present a numerical modeling of the evolution of a pulse in a section of the gain fiber.

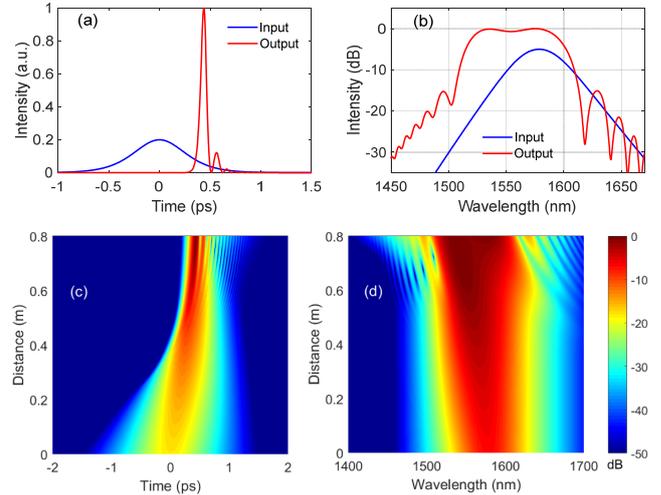

Fig. 3. (a) Input (blue) and output (red) pulse. (b) Input (blue) and output (red) spectra. (c) Waveform evolution in the gain fiber. (d) Spectral evolution in the gain fiber.

To model the pulse evolution in the gain fiber, we use a generalized nonlinear Schrödinger equation with gain as

$$\frac{\partial A(z,t)}{\partial z} - \frac{g}{2} A(z,t) - \sum_{m \geq 2} \frac{i^{m+1} \beta_m(z)}{m!} \frac{\partial^m A(z,t)}{\partial t^m} = i\gamma(z) \times \left(1 + \tau_{shock} \frac{\partial}{\partial t}\right) \times \left[A(z,t) \int_0^\infty R(t') |A(z,t-t')|^2 dt'\right], \quad (1)$$

where $g$ describes the saturated gain coefficient modeled in the frequency domain with a Gaussian profile. Up to the 9-th order dispersion are modeled by $\beta_m$, m=2…9. Terms on the right-hand side of Eq. (1) model the nonlinear effects including self-phase modulation, self-steepening, and Raman scattering response of the fiber. Standard dispersion and nonlinear parameters for single mode fiber are used in the simulation. The gain fiber is assumed to provide a 20 dB small signal gain in 0.8 m length. The gain profile is a Gaussian function with a bandwidth of 5 THz and centered at ~1550 nm. The saturation pulse energy is 1.8 nJ. A positively chirped pulse with a duration of 600 fs and initial bandwidth of ~40 nm centered at ~1580 nm is injected into the gain fiber. The central wavelength of the pulse locates in the L-band of the EDF from the experimental results of the figure-9 seed laser[23]. The pulse is compressed to 65 fs at the output of the gain fiber with some small pulses radiated from the main pulse as shown in Fig. 3(a). The change in the temporal location of the pulse is caused by the dispersion and the wavelength offset of the seed pulse in the simulation window. From Fig. 3(b), a flat-top spectrum with 1-dB bandwidth of ~64 nm, 3-dB bandwidth of ~78 nm and 10-dB bandwidth >100 nm is generated as shown by the red curve. Figures 3(c) and 3(d) demonstrate the temporal and spectral evolutions of the pulse along with the propagation in the gain fiber. Therefore, by engineering the pulse and the fiber amplifier to control the pulse evolution and terminate the evolution at the flat-top point, it is possible to generate a highly coherent signal with a broad flat-top spectrum.

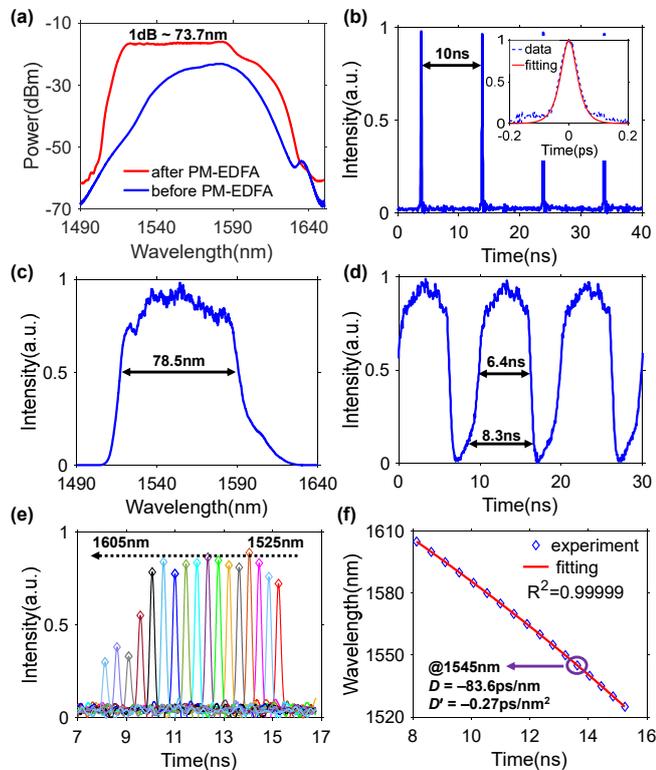

Fig. 4. (a) The output spectra of the mode-locked fiber laser (blue) and the PM-EDFA (red). (b) Pulse train output from the PM-EDFA, the inset shows the autocorrelation trace with a pulse duration of 44 fs. (c) The spectrum on the linear scale after the time stretching in DCF. (d) A pulse train of the 100 MHz swept signal. (e) The waveforms of the swept signal filtered by the TBF when it is tuned from 1525 to 1605 nm. (f) The sweep trace λ(t) obtained by the pulse positions of (e) with parabolic fitting.

As shown by the blue curve in Fig. 4(a), the seed spectrum before the PM-EDFA has a bandwidth of ~40 nm and a center wavelength of 1570 nm [23], which is too narrow for OCT systems. To realize the nonlinear spectral expansion, the dispersion of the PM-EDFA is engineered to eliminate the pulse chirp and support a higher-order soliton evolution of the pulse during the power amplification. The offset between the central wavelengths of the PM-EDFA and the seed spectrum will also help generate a flat and broad spectrum by a stronger amplification to the short wavelength side. The asymmetric amplification, the humps in the spectral profile of higher-order soliton, and the Raman scattering jointly generate a flat broad spectrum. Figure 4(a) shows the broadened spectrum after the PM-EDFA by the red curve. The top-hat spectrum has a 1-dB bandwidth of 73.7 nm. The flatness, bandwidth, and profile of the spectrum qualitatively agree with the spectrum obtained in simulation. To the best of our knowledge, it is the first time that a highly coherent spectrum with such a high flatness over 70 nm is reported. The 3- and 10-dB bandwidths are 79.4 and 106 nm, respectively. The 3-dB bandwidth of the spectrum could be further enhanced to >90 nm by increasing the pump power but the flatness of the pulse top will be degraded by the growing humps on both sides of the spectrum. Along with the spectral broadening, the average power is enhanced to 100 mW and the pulse is simultaneously compressed. The output pulse train has a repetition rate of 100 MHz as shown in Fig. 4(b). From the inset of Fig. 4(b), the full width at half maximum (FWHM) of the pulse is measured to be 44 fs by an auto-correlator (Femtochrome, FR-103HS), which implies a pulse peak power of ~20 kW. The output pulse is very stable since soliton fission for SC generation is not triggered during the nonlinear spectral broadening. The all PM structure of the MLFL and the EDFA greatly enhance the stability of the system [26].

The 100 MHz broadband ultrashort pulse is then used as the seed to obtain swept signal by time stretching technique. The spectrum and temporal waveform of the time stretched swept signal are shown in Figs. 4(c) and 4(d) respectively. The average power after the DCF is 8.3 mW. The output spectrum has a 3-dB bandwidth of 78.5 nm and a 10-dB bandwidth of 105.8 nm, which implies a >100 nm effective sweep range. Meanwhile, slight fluctuations appear on the top of the spectrum owing to the residual nonlinear effect in the DCF. At the output of the DCF, a series of 100 MHz long chirped pulses with an FWHM of 6.4 ns and a duration of 8.3 ns at 10% intensity is obtained as shown in Fig. 4(d). To confirm the wavelength sweep feature of the signal, the signal is filtered by a tunable bandpass filter (TBF, Santec, OTF-930) with a 3-dB linewidth of 0.3 nm and the filter is tuned from 1525 to 1605 nm with a step of 5 nm. The filtered signals are detected by a 3 GHz photodetector and a 4 GHz oscilloscope (Tektronix, CSA7404B) as shown in Fig. 4(e). The pulses for different wavelengths are distributed at different temporal positions, which indicates the sweep feature of the pulses. The sweep trace λ(t) with a parabolic fitting is shown in Fig. 4(f), which presents a quasi-linear positive chirp. From the parabolic fitting with an $R^2$ of 0.99999, the total group velocity dispersion $D$ at 1545 nm is −83.6 ps/nm and the dispersion slope $D'$ is −0.27 ps/nm$^2$, which agree well to the nominal values.

As a result of the high stability of the seed laser, the swept signal is very stable. To quantify the stability, we plot the histogram of interference fringes captured in 998 shots in Fig. 5(a). Figures 5(b)-5(d) are the enlarged views of the histogram at different time positions for better observation. The color indicates the number of events in logarithmic scale. The standard deviation of the fringes normalized to the peak to peak value is 3.1%, which is much lower than that of reported high performance swept source[15] and other tunable filter based swept lasers[8]. The stable interference fringes indicate the high stability of both the waveform and the phase distribution of the sweep signal. This stable swept signal with low timing jitter is crucial for the recalibration to obtain the resampled interference spectrum in the frequency domain.

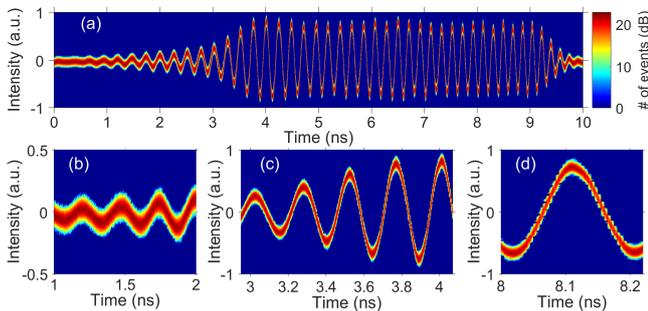

Fig. 5. (a) Histogram of 998 interference fringes. (b), (c), and (d) are the enlarged views of (a).

*B. Point spread function and Sensitivity Roll-off*

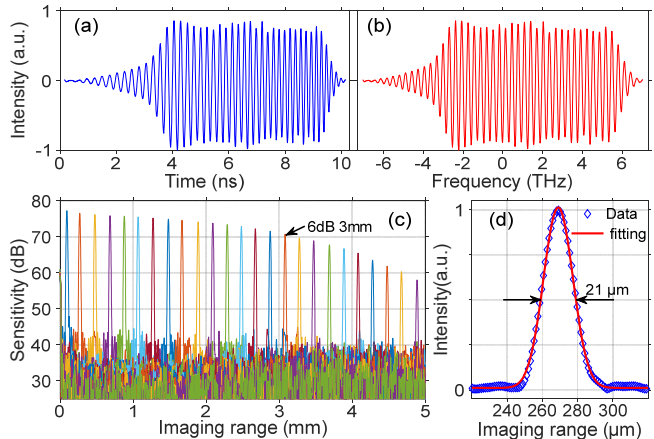

Fig. 6. (a) A typical interference fringes in the time domain. (b) Interference spectrum resampled from (a). (c) Sensitivity roll off characterization with PSFs at different positions. (d) A single PSF shows an axial resolution of 21 μm in air.

A typical single interference fringe captured by the oscilloscope is shown in Fig. 6(a), where a slight chirp in the oscillating frequency can be observed. The slight chirp is caused by the nonzero higher-order dispersion coefficients $\beta_{m>2}$ of the DCF. The temporal signal in Fig. 6(a) is resampled into the frequency domain as shown in Fig. 6(b) with the assistance of the sweep trace $\omega(t)$. The resampled interference spectra in the frequency domain are then used in fast Fourier transform (FFT) with zero paddings to calculate the PSFs. To characterize the performance of the OCT, the VODL in the reference arm is tuned to different positions. In signal processing, the starting points of the swept signal and the period of scan should be chosen accurately to obtain similar narrow PSFs in all scans. If the starting point or the period are not set to the optimal values, the PSFs in scans with alignment error will show degraded profiles with higher sidelobes and larger widths. Such calibration or triggering could be realized with an FBG or other wavelength markers. By FFTs of the resampled interference spectra, a sequence of narrow PSFs locating at different positions is obtained as shown in Fig. 6(c). The sensitivity of the OCT is 77 dB with an average input power of 8.3 mW for the interferometer[5]. The sensitivity could be further enhanced by using low nonlinearity dispersion components such as CFBGs to increase the power. From the PSFs plotted in Fig. 6(c), the 6-dB sensitivity roll-off length is measured to be ~3 mm. The sensitivity roll-off of OCT is determined jointly by the coherence length of the swept source, the sweep slope, and the bandwidth of the detection module [24]. If the coherence length of the swept signal is assumed to be infinite, a simple calculation based on the bandwidths 43 GHz and 59 GHz, respectively, of the BPD and the oscilloscope gives a 6-dB sensitivity roll-off length of ~3.47 mm. The difference between the measured and the predicted values is likely due to the length mismatch between the two fiber tails connecting the last optical coupler to the BPD. A length mismatch of 1 mm could degrade the roll-off length from 3.47 to 3.08 mm, which agrees well with the experimental results.

It should be noted that the flat-top spectrum introduces side lobes on the raw PSFs. By applying a predefined complex window function, the relative intensities of the side lobes of the PSFs are suppressed to –35 dB. In Fig. 6(d), the axial resolution of the OCT is measured to be 21 μm in air with a Gaussian fitting. For comparison, we calculated the best achievable PSF obtained by direct Fourier transform of the spectrum in Fig. 4(c). The PSF also shows distinct sidelobes and an axial resolution of 18 μm. The slight degradation of the axial resolution should come from the sidelobe suppression window function, the residual resample error caused by higher-order dispersion, and the noise introduced in detection. The signal to noise ratio of the PSFs is ~40 dB, which is believed to be determined by the unamplified BPD and the 8 bit A/D conversion of the oscilloscope.

*C. Ultrafast OCT imaging*

The ultrahigh-speed OCT is then used for an imaging test. To demonstrate the ability of ultrahigh-speed detection, we use a rotating encoding disk as the sample as shown in Fig. 7(a). A stainless steel 100-line optical encoding disk with a diameter of 45 mm and a thickness of ~0.2 mm is set on the spindle motor of a hard disk drive with a gap of ~2 mm to the hard disk surface. In detection, the spindle motor is driven at a speed of ~17,000 rpm. The tangential line speeds of the edges of the encoding disk and the 70 mm hard disk are 40 m/s and 62 m/s, respectively. As shown by the red arrow in Fig. 7(a), the sample beam is normally incident on the disks with an adjustable fiber collimator. The lights reflected alternatively by the rotating encoding disk and hard disk are collected by the collimator and return to the fiber for detection.

An uninterrupted sequence of A-scan data with a duration of 4 ms is captured by the oscilloscope. By resolving all the A-scan data and applying FFTs, the tomographic image of the profile along the periphery of the encoding disk with a perimeter of ~126 mm is shown in Fig. 7(b). The two rows of dots near the top and bottom of the tomographic image show the reflection of the encoding disk and the hard disk, respectively. For better observation, the reflectivity profiles are shown in logarithmic scale in a range of 30 dB. The measured distance between the reflective surfaces of the two disks is ~2.2 mm with slight variations on different azimuthal angles. Such depth variations are caused by the imbalance of the fastening force of the 6 holding screws on the spindle motor and the low flatness of the thin steel encoding disk. The detail of the tomography in the azimuthal angle range of 199~211° with a dimension of ~4.2 mm (L) × 2.5 mm (D) is shown in Fig. 7(c). In the tomographic image, the reflectivity of the encoding disk is stronger than that of the hard disk because of the sensitivity roll-off since the hard disk is far from the zero position. Figures 7(d) and 7(e) are the zoom-in views of the small areas in Fig. 4(b) indicated by the two red rectangles in the range of 11~21° and 265~275°, both with a dimension of ~3.5 mm (L) × 0.15 mm (D). From Figs. 7(c)-7(e), the patterns of the hard disk are very uniform but fluctuations appear on the patterns of encoding disk. The higher uniformity demonstrates that the state-of-art precisely fabricated hard disk has a much smoother surface when compared with the steel encoding disk which is massively produced by punches.

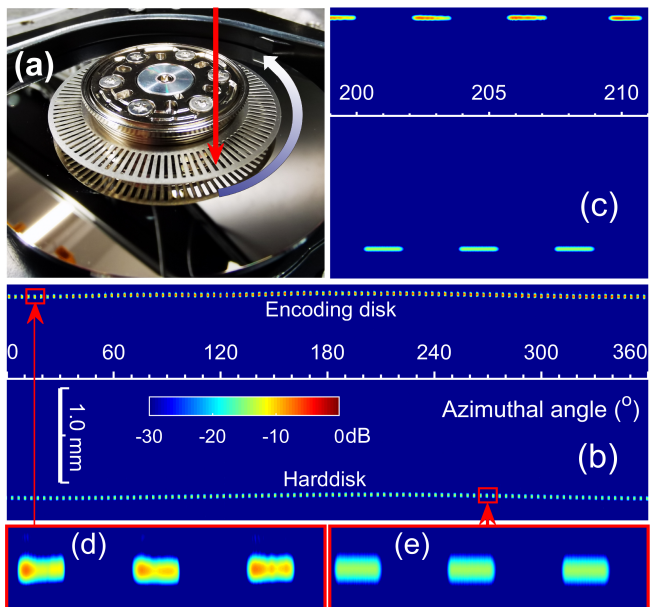

Fig. 7. (a) Photograph of the encoding disk and hard disk. (b) Tomographic image of the profile of the rotating disks. (c) Detail of the tomographic image in the range of 199~211° with a dimension of ~4.2 mm × 2.5 mm. (d) and (e) are the zoom-in views of the small regions indicated by the red rectangles in (b) with a dimension of ~3.5 mm × 0.15 mm. The color bar applies to (b)-(e).

## IV. Conclusion

In conclusion, we demonstrate an ultrafast SS-OCT with an A-scan rate of 100 MHz based on a swept source by time stretching of broadband femtosecond pulses from a mode-locked fiber laser. The swept source shows an effective sweep range over 100 nm and an ultra-flat spectrum with a 1-dB bandwidth of 73.7 nm. The SS-OCT has an axial resolution of 21 μm in air and a 6-dB sensitivity roll-off length of 3 mm. A tomographic image of an encoding disk and a hard disk rotating at 17,000 rpm is captured by the SS-OCT. To the best of our knowledge, it is the first time that an SS-OCT demonstrates an A-Scan rate of 100 MHz and a sweep range of 100 nm simultaneously. The ultrafast SS-OCT enables the tomographic characterization of high-speed moving targets, which could also be used to enhance the frame rate of video-rate volumetric imaging with SS-OCT.

**Dongmei Huang** received her B.Sc degree in 2014 from Huazhong University of Science and Technology, and obtained her M.Sc degree in 2017 from Chongqing University, and obtained her Ph.D degree in 2020 from the Hong Kong Polytechnic University, China.

She is currently a postdoctoral fellow at Photonics Research Centre of the Hong Kong Polytechnic University. Her research interests include wavelength swept lasers and its applications in optical coherence tomography and optical sensing systems, nonlinear microresonators.

**Feng Li** received the B.Sc and Ph.D degrees from the University of Science and Technology of China, Hefei, China, in 2001 and 2006, respectively.

After that, he joined The Hong Kong Polytechnic University, Kowloon, Hong Kong, as a Postdoctoral Fellow. He is currently a Research Assistant Professor in The Hong Kong Polytechnic University. His research interests include fiber lasers including mode locked fiber laser, multi-wavelength fiber laser, wavelength swept laser and its application in bio-photonics; nonlinear optics in optical fiber, micro-resonator/waveguides, including soliton propagation and supercontinuum generation; optoelectronic devices.

**Zihao Cheng** received the B.Sc and M.Sc degrees from Peking University, Beijing, China, in 2014 and 2017, respectively. He is currently working toward the Ph.D degree at The Hong Kong Polytechnic University, Kowloon, Hong Kong.

His research interests include integrated optics, silicon photonics, and nonlinear microresonators.

**Xinhuan Feng** received the B.Sc degree at physics department of Nankai University in 1995, and obtained her M.Sc and Ph.D degrees respectively in 1998 and 2005 at Institute of Modern Optics, Nankai University, China.

From 2005 to 2008, she worked as a postdoctoral fellow at Photonics Research Centre of the Hong Kong Polytechnic University. Since March 2009, she has been with the Institute of Photonics Technology, Jinan University, China. Her research interests include various fiber active and passive devices and their applications, and microwave photonic signal processing

**P. K. A. Wai** received the B.Sc (Hons.) degree from the University of Hong Kong, Hong Kong, Hong Kong, in 1981, and the M.Sc and Ph.D degrees from the University of Maryland, College Park, MD, USA, in 1985 and 1988, respectively.

In 1988, he joined Science Applications International Corporation, McLean, VA, USA, where he was a Research Scientist involved with the Tethered Satellite System project. In 1990, he became a Research Associate in the Department of Physics, University of Maryland, College Park, and the Department of Electrical Engineering, University of Maryland, Baltimore County, MD. In 1996, he joined the Department of Electronic and Information Engineering, The Hong Kong Polytechnic University, Kowloon, Hong Kong. He became the Chair Professor of optical communications in 2005. His research interests include soliton, fiber lasers, modeling and simulations of optical devices, long-haul optical fiber communications, all-optical packet switching, and network theories. He is an active contributor to the field of photonics and optical communications, having authored or coauthored over 500 international refereed publications.

Prof. Wai is an OSA Fellow, IEEE Fellow and Fellow of the Hong Kong Academy of Engineering Sciences.